# Assessment of Financial Potential as a Determinant of Enterprise Development

Dmytro Zherlitsyn
*Finance and Audit Department*
*Zaporizhzhya Institute of Economics and Information Technologies*
Zaporizhzhia, Ukraine
http://orcid.org/0000-0002-2331-8690

Stanislav Levytskyi
*Department of Economic Cybernetics*
*Zaporizhzhya Institute of Economics and Information Technologies*
Zaporizhzhia, Ukraine
http://orcid.org/0000-0001-6014-1276

Denys Mykhailyk
*Department of Inetnational Economic Relations*
*Zaporizhzhya Institute of Economics and Information Technologies*
Zaporizhzhia, Ukraine
http://orcid.org/0000-0003-2425-0437

Victoriia Ogloblina
*Finance and Audit Department*
*Zaporizhzhya Institute of Economics and Information Technologies*
Zaporizhzhia, Ukraine
http://orcid.org/0000-0001-6627-0255

*Abstract*—**Financial potential is an important part of enterprise activities. The technique of the enterprise's financial potential assessment is offered in the paper. It is presented by particular stages, where each stage is related to a certain task. The characteristics of the company's financial potential, based on the analysis of the related literature, are determined. The implementation of each task is carried out. Thus, the study proposes a mechanism for managing the financial potential of enterprises, which allows to emphasize the elements that can be useful for economic development. It is based on the general strategic principles of the enterprise management. The study results can be used to assess enterprise purposes and develop the formation goals of its financial potential. It can also help to forecast and separate main directions of accumulation, formation, and distribution of financial resources. It should be noted, that analysis and control over the financial potential formation strategy, as well as the use of analysis results for specifying the strategic directions of the enterprise development, are of high importance. Therefore, the management of the financial potential is a system of rational management of business financing, which includes the formation of financial relations, emerging as a result of finance resources flow.**

*Keywords—assessment, financial potential, development, determinant, enterprise*

## I. Introduction

There are different approaches for determining the structure of the enterprise's potential in the latest economic researches. Many of the scientists consider the resource structure of the potential but everyone allocates a different number of components [1]. Some researchers include in the composition of the potential only the means of work, others – the means of work and workforce, the third – the means of work, workforce, and natural resources used in the production process, the fourth – means of work, workforce, and objects of work.

## II. The Latest Features of Financial Potential Assessment

There are many approaches for defining strategic (economic) potential both in the latest economic studies. L. Benati [11], A. Bilandzic [12], M. Jeger, N. Sarlija [14], P. V. Puzyryova [15] and N. Trusova [16] define that strategic potential for regions, corporates or SMEs combine all available material and intellectual resources. Therefore, the strategic potential is a system of both existing and possible knowledge, skills, and abilities of employees, as well as a set of economy resources. They allow to develop and implement a strategy that will ensure the effective functioning, development, and preservation of the organization's competitive position in the market.

O. Fedonin [2] defines the following characteristics of the company's strategic potential, that help to determine its potential capabilities.

- Allows conducting a macroeconomic analysis of the external environment of the enterprise.

- Allows building forecasts of changes in the structure and volume of demand for products or services of the enterprise.

- Allows analyzing, as well as forecasting the conjuncture of capital and resource markets.

- Provides the ability to develop and implement effective strategies for the interaction of the enterprise with different markets to attract necessary resources.

- Ensures the enterprise's sustainability to negative changes in the external environment by developing and implementing effective protective strategies.

- Ensures the ability to effectively use of investment opportunities to develop constituent capacities.

There are many other approaches to determine the structure of strategic potential in the latest economic researches. Some of them assumes strategic potential as one of the final products that ensure the achievement of the enterprise's objectives. The input of strategic potential comes from human resources, raw materials, information, and financial resources. The results or the outputs of strategic potential are the products or services produced, as well as a set of activities that help the company to achieve its goals [1], [2], [14], [15], [16].

The structure of the strategic potential is determined by the basic elements of the enterprise activities and the quality of the personnel. There are following key elements [2]: features and capacities of technologies, equipment, buildings, and structures; features and capacities of communication systems for the transferring and processing of information; the organizational structure of the enterprise, i.e. distribution of job tasks and responsibilities as for a separate employee or for groups of workers; internal processes; organizational culture, such as norms, rules, values, and traditions, laid on the basis of the organizational behavior of the enterprise.

The quality of the personnel can be evaluated by following [4]: professional qualifications and competence of employees; adaptation to changes; ability to solve problems at the strategic level. Thus, according to this approach, the structure of strategic potential reflects the direction of management to ensure the long-term viability of the organization within constantly changing conditions. The early reviewed structure of the enterprise's potential clearly demonstrates the place and role of the strategic capacity. The strategic potential exists when there is an interaction of the main potentials of the enterprise. They include financial, investment, economic, marketing, logistics, production, resource, intellectual, scientific and technical, innovative, managerial potential, as well as the individual potential of engineering and technical staff.

According to R. Kaplan and D. Norton [5], the description of the balanced scorecard is based on the "bottom-up" principle; therefore, considering the structure of the strategic potential, it is necessary to start from the perspective of Training/Development. This kind of potential includes the managerial potential and individual potential of engineering and technical staff [3]. The second component of the Training/Development perspective is the innovation potential, where the scientific personal and technical staff are the components of it.

The latest studies show that the rationale for the concepts of "intellectual potential" and "human potential" is equal. Yu. Lysenko [3], M. Olisckevich and I. Loukianenko [4], I. Ansoff [6] define that there is no unambiguous approach to the definition of human potential. The authors prove that the human potential is a combination of both the physical and intellectual abilities of the employee, which can be used to achieve strategic goals and solve specific tasks of the enterprise. The importance of management capacity at the enterprise is high. It assumes the ability to see the course of actions in advance, as well as to set strategic goals and develop tactics for achieving them, especially in problematic situations. And, of cause, it is one of the main keys to successful operation of the enterprise.

The meaning of "managerial potential" is a fairly new and, therefore, there is still no generally accepted interpretation of it. However, it should be noted that I. Ansoff [6] in his works suggests interpreting it as the volume of work that management can undertake under the management potential. A. Fedonin [2] defines managerial potential as a set of knowledge, skills, and abilities of managers according to the structure of the enterprise, which they apply to create the necessary conditions for the operational and development processes [1].

According to L. Prokopyshyn [7], management potential is the knowledge, skills of the personnel that can be applied to make the most effective decisions, to ensure the successful operation of the enterprise, and to achieve the set of goals. The meaning of "the potential of engineering and technical personnel" is not widely used in the modern scientific studies. So, it is necessary to define the term "engineering personnel and technical staff" [7] According to the analysis of various sources of personnel management, it can be concluded that the engineering staff is a group of employees who organize and manage the production process at the enterprise.

Consequently, the potential of engineering and technical personnel is understood as the hidden knowledge and skills of employees that they do not currently use but can implement while solving tasks related to the production process of the enterprise. Touching upon the topic of personnel management from the position of the enterprise's potential, it is necessary to consider the formation of the personnel reserve at present time.

A personnel reserve is a group of employees at the enterprise who have the ability to lead activities and their further professional development can be used to replace certain positions. The next part of the Training/Development perspective is the innovative potential of the enterprise. Some scientists do not share the notion of "innovation potential" with components and consider the term in a broad sense, as the ability and readiness of an enterprise to develop and implement innovations. In this paper, on the contrary, the scientific and technical potential has been identified within the structure of innovative potential. This is due to the necessity to isolate the work that can be done with the help of engineering and technical personnel to build productive capacity.

The perspective of Business Processes block includes the production and resource potential that directly affect the enterprise. The production potential is a system that allows transforming the resource potential into finished products (goods, services).

The specific of the Market/Clients block should be noted. G. Bagiev [8] defines that the marketing potential is "…an integral part of the enterprise's potential, ensuring its constant competitiveness, economic, and social conjuncture of the goods in the market through effective marketing activities in the field of demand research, commodity, price, communication, distribution, marketing policy, organization of strategic planning and control over the behavior of goods, competitors, consumers, and the enterprise itself in the market" [8]. Lots of managers take into account only a part of its customers and adapts resources to the specific requirements of customers. So the task of marketing potential is to find ways to establish a balance between own resources and market potential.

### III. Defining The Financial Potential Of The Enterprise

In the structure of strategic potential, the "Finance" perspective is represented by economic, investment, and financial potential. Investment potential is the implementation of the developed investment policy principles into practice. It determines the properties and dynamism of investment activity. The main condition for the justification of the term "investment potential" is the need for the enterprise to have the knowledge and the possibility of attracting investment resources together with the organization's ability to increase the market value of the business through the implementation of investment projects. Thus, the investment potential is based on the knowledge that allows adapting the intellectual and production potential for increasing the financial potential. So, the investment potential is an opportunity for an enterprise to invest in its own development.

The authors have argued that the formation and existence of the enterprise investment potential is aimed to solve the following problems:

- continuous development of the organization and production through the search, selection, development, as well as the implementation of innovative ideas;
- formation of the base of innovative proposals and ways for their implementation;
- organization of the processes of problems' identification, that may impede the development of the enterprise, as well as the development of measures to address them;
- encouragement of the staff to develop innovative ideas, creating a climate of innovation.

The financial potential reflects potential investment opportunities and financial indicators, such as profitability, liquidity, solvency. The competitiveness of this type of potential is expressed in the firm's sustainable solvency, as well as in the availability of sufficient working capital, which, with the help of a clear organization of calculations, is appropriately and effectively used in economic activities [9]. For example, E.I. Altman, T.K. Baidya, L.M. Riberodias [10], A. I. Borodin [13] include a lot of different processes and indicators to the enterprise financial potential. There are rates of growth, total debt, equity, debt and so on.

The authors have suggested that the structure of the enterprise's potential requires optimization of business processes and synergy effect assessment. So, those processes should be carried out on the following stages:

Step 1. Formulate the strategic, tactical, and current goals for the entire enterprise, and for its units and individual activities.

Stage 2. Define a set of strategic resources for each goal. The necessary coordination of this process is optimizing of the structure of the enterprise's potential with the first stage.

Stage 3. Propose an alternative variants of strategic resource sets for the purposes of the enterprise, their evaluation, the possibility of combinations of several sets. Formulating final conclusions on the choice of resources.

Stage 4. Rationally distribute the selected resources and determine their profitable direction to ensure the high competitiveness of the enterprise's potential.

Step 5. Evaluate the results obtained after the first four stages of optimizing the structure of the enterprise's potential.

The necessity to form and implement a financial strategy as a necessary element of the financial potential management of businesses is determined by the deepening of market reforms, integration processes' development, and growth of volatility factors in the outer financial environment. The financial strategy here means a flexible general model of the financial system of the enterprise development, which ensures its sustainability and adaptation to changing economic situation, in other words, the formation of stable financial potential happens. Management of financial potential is a system of rational management of business financing, which includes the formation of financial relations emerging as a result of finance resources flow.

Financial potential of stability and well-being characterizes the possibility of the organization and the enterprise's current activities results, while the financial potential of development suggests the possibility of further development. Thus, financial potential of enterprises' development is the sum-total of all existing potential resources, including financial ones, which can ensure definite strategic goals achieved in current and long-term perspective, considering outer factors influence.

Growth of the financial potential of modern business is accompanied by variability, unpredictability, and strengthening of economic globalization, and depends on a considerable extent of their financial stability. There are next conditions for the possible existence of financial potential [17].

- Availability of own capital in the value, which is enough for business activities and financial stability.
- Opportunities for gaining loans for the business development.
- Cost-effectiveness of investment.

It should be noted that the financial potential is reflected to total cash flow of business, which influences the state of potential and its development trends. The essence of financial potential management bases on the effectiveness of using the mechanism of financial strategy management for business's goals achievement. Basic categories of the theory and practice of evaluation, formation, and management of the enterprise are financial resources, cash flow, profit, current assets, and net assets.

The authors conclude that there are few main segments that make the enterprise's management system of financial potential more efficient.

- Current liabilities management.
- Paying capacity, liquidity, financial stability management.
- Capital management.
- Enterprise's investments management.
- Cash flow management.
- Long-term loans management.
- Indirect investments management and business evaluation.
- Pricing and enterprise's capital structure management.
- Entrepreneurial risks management.

The specifics of enterprise's financial potential management are defined by next elements [11].

- List of financial resources.
- Necessity of rational distribution of financial resources, which allows reaching and sustaining balance, financial stability of the enterprise, and gaining profit along with financial backup for economic and social tasks.
- Competition influence accounting.
- Need for gaining of financial resources from different sources.
- Image of management object based on cash flow sizes and payment terms.
- Use of management influence on financial resources.
- Financial potential structural reforms regulation.
- Development and decision making after reaching compromise on profitableness, reliability, and liquidity of enterprises' capital.

Therefore, the enterprise's financial potential management is a complex of stages, financial methods, financial tools, and organizational support, which are optimally coordinated. There are following instruments of their coordination [1].

- Financial methods, which include forecasting, planning, insurance, investment, loaning.
- Financial leverages applied to identify the payment conditions, prices, types of loans, and interests.
- Financial tools, including the securities, stock options plan, funds structure, futures and forwards contracts and so on.

Therefore, the enterprise's financial potential management strategy is a system of coinciding goals and actions, the development and realization of these goals and actions, which are based on financial management. The given system allows managing the plans of gaining and distributing of financial resources. Development of the enterprise's financial potential strategy should be based on the effective use of the market rules, which main goal is securing cash flow among their owners. The enterprise's financial potential strategy development includes assessment of the set of financial tools, methods, laws, etc. However, the market tools are evolving and renovating constantly, and this fact makes continuous research of market development trend, along with in time decision-making on profitable economic strategies, the main factors of enterprise's effective activity.

The authors argued that one of the main stages of forming a strategy for managing the financial potential of enterprise is a forecast of financial potential. The financial potential forecast is a formation of a financial capacity system, as well as a selection of the most effective ways to optimize it. During the financial forecast process, the development of the general financial growth concept and the enterprise's financial policy regarding certain aspects of its activity comes out. It worth mentioning that financial forecast cannot be called accurate since a lot of the important factors can lead to the discrepancy between forecast results and reality. Nevertheless, the possibility of mistake is not the reason to abandon forecasting.

It is possible to set some of the financial forecast aspects. They are an enterprise's resource-based forecast, including labor forces, financial and material resources; a financial situation forecast made by evaluation of financial balance, a financial policy forecast.

For the financial potential forecast to be more accurate, the information used should be of high authenticity and also well-structured and given in a full measure. In addition, there should be a possibility to collate information upon quality and quantity indicators. One of the specific characteristics of the financial potential forecast is that the enterprise has a certain level of interdependence and sluggishness, where the sluggishness level is the dependence on of any index in the present moment from the previous period. However, the forecast index includes the realization of the accurately determined model, correlating result with factors, which influenced the given forecast.

Thus, financial forecasting sets the task of the development of financial potential management strategy optimization model, along with the research of important factors. It can be separated into internal and external factors, which promote the development of financial potential elements and effect their balance and effectiveness.

The study results define the main management systems, which in prospect allow efficiently realizing the enterprise's financial potential management strategy.

1. The management of enterprise's development stability, liquidity, and financial reliability. The given system means the management of accounts receivable, short-term marketable securities, short-term credit debts, funds, and stockpiles.

2. The management of enterprise's investment activity that means carrying out of the comparative analysis of different investment projects efficiency, along with setting out the value of cash flows with the allowance for the influence of the next factors: risk and inflation, time, and setting of criteria for the financial decision-making.

3. The management of enterprise's financial sources by ensuring the stable growth of its own capital, based on increasing of nominal capital by issuing capital stocks, usage of effective dividend policy and increasing profits. Also, management of such system can be realized by long-term fund raising with the help of leasing, long-term credits and corporate bond issuance.

4. The management of the enterprise's financial stability by estimating of its value, formation of an optimal capital structure, estimation of the enterprise's total value, and setting of optimal correlation between leveraged investments and owned assets.

5. Control over cash flow.

6. Maximization of enterprise's profits with allowance for financial risk and minimization of financial risk level considering the level of profits needed.

It should be noted that one of the most important stages of the formation mechanism of the enterprise's financial potential management strategy is the analysis and evaluation of financial risks. Recognition of financial risks includes identification of every possible risk connected to the certain corporate operation. In addition, it is necessary to recognize risks that depend on the enterprise itself and on its external environment, defined by macroeconomic activity.

When identifying risk factors, it is usable to divide them into internal and external, where external financial risks affect economic and market factors. Fist group is the economic risks factors. There are next indicators: rise in the inflation rate, instability, and imperfection of tax laws, the decrease in the state manufacture volume, the decrease in the income level and purchasing capacity of the population, the delay of payment transaction etc. Market factors are the decrease of market demand, lowering home market capacity etc. Internal financial risks of the enterprise are influenced by investment, financial, and production commercial factors.

## IV. Conclusion

To sum up the offered results, it should be pointed out that the mechanism of the enterprise's financial potential management strategy formation is presented by correlating stages, where each stage is oriented to a certain task. The authors obtained the result where the main stages of the mechanism of financial potential management formation include next elements.

- Analysis of the enterprise's strategy and allocation of the elements, which help to formulate the strategy of financial potential management.

- Development of goals, which consists of analysis and forecast of economic conditions of the environment; analysis of the enterprise's internal environment; identification of strengths and limitations.

- Leading policy of adaptation to environment conditions.

- Searching for new sources for getting loans, and definition of financial resources' accumulation, formation, and distribution strategies.

- Definition of financial potential management strategies

- Financial tools, tax planning, organizational and legal execution, planning and control over cash flows and resources, along with analysis and evaluation of financial risks.

- Development and realization of financial potential management strategy.

- Financial potential management analysis and control.

- Using results of the analysis for the specification of the enterprise's development strategy.

These elements help to form financial potential management strategy and define the enterprise's competitiveness. Execution of these elements along with the enterprise's management system allows realizing the main goals of the enterprise.

Therefore, the study results have shown the mechanism of forming the enterprise's financial potential management, based on enterprise's general strategy and allowing to emphasize the elements useful in the creation of financial potential. It worth mentioning that analysis and control over financial potential formation strategy are both important, as well as the usage of analysis results for the concretization of main strategic directions of the enterprise's development.

Implementation of those results will allow developing financial potential formation goals, building forecast, separating the main directions of accumulation, formation, and distribution of financial resources as well as using of assessment of financial potential as enterprises development determinant.